\documentclass[onecolumn,showpacs,floatfix,11pt,nofootinbib]{revtex4}
\usepackage{amssymb,amsmath}
\usepackage[dvips]{graphics,color}
\usepackage{graphicx}
\usepackage{float}

\newcommand{\ba}{\begin{eqnarray}}
\newcommand{\ea}{\end{eqnarray}}
\newcommand{\bege}{\begin{equation}}
\newcommand{\enge}{\end{equation}}
\newcommand{\beq}{\begin{eqnarray}}
\newcommand{\benu}{\begin{enumerate}}
\newcommand{\enu}{\end{enumerate}}
\newcommand{\eeq}{\end{eqnarray}}

\input{tcilatex}

\begin{document}

\title{Friedmann-Robertson-Walker Braneworlds}
\author{P. Michel L. T. da Silva}
\email{pmichel@fc.unesp.br}
\affiliation{Departamento de F\'{\i}sica e Qu\'{\i}mica, Universidade Estadual Paulista,
Av. Dr. Ariberto Pereira da Cunha, 333, Guaratinguet\'{a}, SP, Brazil.}
\author{A. de Souza Dutra}
\email{dutra@feg.unesp.br}
\affiliation{Departamento de F\'{\i}sica e Qu\'{\i}mica, Universidade Estadual Paulista,
Av. Dr. Ariberto Pereira da Cunha, 333, Guaratinguet\'{a}, SP, Brazil.}
\author{J. M. Hoff da Silva}
\email{hoff@feg.unesp.br}
\affiliation{Departamento de F\'{\i}sica e Qu\'{\i}mica, Universidade Estadual Paulista,
Av. Dr. Ariberto Pereira da Cunha, 333, Guaratinguet\'{a}, SP, Brazil.}

\begin{abstract}
We study the cosmological evolution with nonsingular branes generated by a
bulk scalar field coupled to gravity. The specific setup investigated leads
to branes with a time-dependent warp factor. We calculate the effective
Hubble parameter and the effective scale factor for the FRW branes obtained
solutions. The spatially dependent branes solutions also were found.
\end{abstract}

\maketitle

\section{Introduction}

Braneworld models were introduced as a genuine branch of research in high
energy physics since the outstanding works presented in Refs. \cite%
{RS1,RS2,Gob}. After those works, several authors tried to construct models
compatible with the Friedmann-Robertson-Walker cosmology. In fact, the
modeling of realistic braneworld scenarios from the large scale physics
point of view must include, at least in some level, the description encoded
in the cosmological standard model. The attempts put forward to this program
can be roughly separated into two categories: one dealing with infinitely
thin branes, in which the extra dimensional effects enters as corrections to
the Einstein equations via the presence of the Weyl tensor (the so-called
`dark fluid') and quadratic contributions to the stress tensor, and another
one, whose modeling arrive at thick branes described by one or more bulk
scalar fields. In the former approach, the corrections to the gravitational
equations comes from the Gauss-Codazzi procedure and, potentially, all the
relevant aspects of the four-dimensional cosmology are revisited. It turns
out, however, that an infinitely thin brane seems to be only an
approximation of the more realistic case, at best. In fact, by keeping in
mind the simple fact that at very short scales a classic gravitational
theory must be replaced by its quantum counterpart (whatever it is), it is
mandatory some thickness to the brane itself.

Unfortunately, in the thick brane context it is not possible, at the best of
our knowledge, to apply the Gauss-Codazzi formalism. The reason is that it
is not clear what (if any) are the Israel-Darmois junction conditions in a
thick brane context. The junction conditions are at the heart of the
projection procedure, and its lack makes the whole program fall apart. In
this vein, the investigation of a five-dimensional thick braneworld setup
whose four-dimensional part describes a Friedmann-Robertson-Walker (FRW)
universe is, indeed, in order. There are, nevertheless, only a few works
addressing this crucial point. The cumbersome algebraic task inevitably
present in this endeavor can be attributed as a cause for such.

In this paper we address ourselves to this task, starting from a five
dimensional scalar field whose solution describes a FRW brane, i. e., a
braneworld whose four-dimensional part is given by a FRW universe. The main
braneworld characteristic, the warp factor, is also take into account in the
solution.

We organize this paper as follows. In Section II, we present the
mathematical preliminaries supporting the metric ansatz with the
four-dimensional background metric given by the Friedmann-Lema\^{\i}%
tre-Robertson-Walker. Then we derive gravitational equations and the
expressions for the scalar field and its potential. In Section III, we find
the time-dependent FRW brane solutions. We solve the field equations for two
cases with respect to spatial curvature, i.e, $k=0$ and $k\neq 0$. Going
further we determine the spatial part of the set of equations in Section IV.
In Section V we present the effective Hubble parameter as well as the
effective scale factor for all the possibilities found previously. In the
final section we conclude.

\section{Field Equations}

The assumption of isotropy and homogeneity implies the large scale geometry
described by a metric of the form%
\begin{equation}
ds^{2}=-dt^{2}+a^{2}(t)\left[ \dfrac{dr^{2}}{1-kr^{2}}+r^{2}(d\theta
^{2}+sen^{2}\theta d\phi ^{2})\right] ,  \label{FRWL}
\end{equation}%
in a synchronized coordinate system (a suitable set of coordinates called
comoving coordinates). The comoving observers, also called Hubble observers,
are the ones located at spacelike hypersurfaces accompanying the cosmic
fluid, which is at rest with respect to such hypersurfaces. Here $a(t)$ is
an arbitrary function of the cosmic time called scale factor and $k=0,$ $\pm
1$, denotes the spatial curvature of the universe for Minkowski, Riemann and
Lobachevsky geometry, respectively.

Let us consider 5D spacetimes for which the metric takes the following form%
\begin{equation}
ds^{2}=a^{2}(t,y)\left\{ -dt^{2}+u^{2}(t)\left[ \dfrac{dr^{2}}{1-kr^{2}}%
+r^{2}(d\theta ^{2}+sen^{2}\theta d\phi ^{2})\right] \right\}
+b^{2}(t,y)dy^{2},  \label{1}
\end{equation}%
where the background metric in $4D$ is given by the Friedmann-Lema\^{\i}%
tre-Robertson-Walker line element (\ref{FRWL}). The metric signature is
given by $(-++++)$. The function $a(t,y)$ is an warp factor with time and
extra dimension dependence, while $u(t)$ performs the usual scale factor for
an homogeneous and isotropic universe. The function $b(t,y)$ shows the
dynamics of the extra dimension at different times and positions in the bulk.%
\textit{\ }

Let us consider the 5D action in the presence of a bulk scalar field with
the potential $V(\phi )$ minimally coupled to the gravitational sector%
\begin{equation}
S=\int d^{5}x\sqrt{-g}\left\{ 2M^{3}R-\frac{1}{2}g^{MN}\nabla _{M}\phi
\nabla _{N}\phi -V(\phi )\right\} ,  \label{2}
\end{equation}%
where $M$ is the Planck mass and $R$ is the five-dimensional Ricci scalar.
In general we suppose that the scalar field $\phi $ depends only on time and
the extra dimension $y$.

The Einstein equations read 
\begin{equation}
R_{MN}-\frac{1}{2}g_{MN}R=\frac{1}{4M^{3}}T_{MN},  \label{3}
\end{equation}%
and the energy momentum-tensor $T_{MN}$ for the scalar field $\phi (t,y)$ is%
\begin{equation}
T_{MN}=\nabla _{M}\phi \nabla _{N}\phi -g_{MN}\left( \frac{1}{2}g^{AB}\nabla
_{A}\phi \nabla _{B}\phi +V(\phi )\right) .  \label{4}
\end{equation}%
The time-time component of the field equations for the space-time under
consideration is given by 
\begin{gather}
\text{ \ \ \ \ }3\left[ \dfrac{1}{a^{2}}\dfrac{\overset{.}{a}^{2}}{a^{2}}-%
\dfrac{1}{b^{2}}\left( \dfrac{a^{\prime \prime }}{a}+\dfrac{a^{\prime 2}}{%
a^{2}}\right) +\dfrac{1}{b^{2}}\dfrac{a^{\prime }}{a}\dfrac{b^{\prime }}{b}+%
\dfrac{1}{a^{2}}\left( 2\dfrac{\overset{.}{a}}{a}\dfrac{\overset{.}{u}}{u}+%
\dfrac{\overset{.}{a}}{a}\dfrac{\overset{.}{b}}{b}+\dfrac{\overset{.}{u}^{2}%
}{u^{2}}+\dfrac{\overset{.}{b}}{b}\dfrac{\overset{.}{u}}{u}+\dfrac{k}{u^{2}}%
\right) \right]  \notag \\
=\dfrac{1}{2b^{2}}\phi ^{\prime 2}+\dfrac{1}{2a^{2}}\overset{.}{\phi ^{2}}%
+V(\phi ),  \label{6}
\end{gather}%
whilest the the space components give%
\begin{gather}
\text{\ \ \ }\dfrac{1}{a^{2}}\left[ 2\dfrac{\overset{..}{a}}{a}-\dfrac{%
\overset{.}{a}^{2}}{a^{2}}+\dfrac{\overset{.}{a}}{a}\dfrac{\overset{.}{b}}{b}%
+4\dfrac{\overset{.}{a}}{a}\dfrac{\overset{.}{u}}{u}+2\dfrac{\overset{.}{b}}{%
b}\dfrac{\overset{.}{u}}{u}+\dfrac{\overset{..}{b}}{b}+2\dfrac{\overset{..}{u%
}}{u}+\dfrac{\overset{.}{u}^{2}}{u^{2}}+\dfrac{k}{u^{2}}\right]  \notag \\
-\dfrac{3}{b^{2}}\left[ \dfrac{a^{\prime \prime }}{a}+\dfrac{a^{\prime 2}}{%
a^{2}}-\dfrac{a^{\prime }}{a}\dfrac{b^{^{\prime }}}{b}\right] =\dfrac{1}{%
2b^{2}}\phi ^{\prime 2}-\dfrac{1}{2a^{2}}\overset{.}{\phi ^{2}}+V(\phi ).
\label{7}
\end{gather}%
The extra dimensional part contributes with 
\begin{equation}
\text{ \ \ \ }\left[ \dfrac{a^{\prime }}{a}\dfrac{\overset{.}{a}}{a}+\dfrac{%
a^{\prime }}{a}\dfrac{\overset{.}{b}}{b}-\dfrac{\overset{.}{a}^{\prime }}{a}%
\right] =\dfrac{1}{3}\overset{.}{\phi }\phi ^{\prime },  \label{8}
\end{equation}%
\begin{gather}
\text{\ \ \ \ }3\left[ 2\dfrac{a^{\prime 2}}{a^{2}}\dfrac{1}{b^{2}}-\dfrac{%
\overset{..}{a}}{a}\dfrac{1}{a^{2}}-3\dfrac{\overset{.}{a}}{a}\dfrac{\overset%
{.}{u}}{u}\dfrac{1}{a^{2}}-\dfrac{\overset{.}{u}^{2}}{u^{2}}\dfrac{1}{a^{2}}-%
\dfrac{\overset{..}{u}}{u}\dfrac{1}{a^{2}}-\dfrac{k}{a^{2}u^{2}}\right] 
\notag \\
=\dfrac{1}{2a^{2}}\overset{.}{\phi ^{2}}+\dfrac{1}{2b^{2}}\phi ^{\prime
2}-V(\phi ).  \label{9}
\end{gather}%
Finally, the scalar equation of motion reads 
\begin{equation}
\dfrac{1}{b^{2}}\phi ^{\prime \prime }-\dfrac{1}{a^{2}}\overset{..}{\phi }+%
\dfrac{4}{b^{2}}\dfrac{a^{\prime }}{a}\phi ^{\prime }-\dfrac{2}{a^{2}}\dfrac{%
\overset{.}{a}}{a}\overset{.}{\phi }-\dfrac{1}{a^{2}}\dfrac{\overset{.}{b}}{b%
}\overset{.}{\phi }-\dfrac{3}{a^{2}}\dfrac{\overset{.}{u}}{u}\overset{.}{%
\phi }-\dfrac{1}{b^{2}}\dfrac{b^{^{\prime }}}{b}\phi ^{\prime }-\dfrac{dV}{%
d\phi }=0,  \label{10}
\end{equation}%
where a dot denotes a derivative with respect to $t$, and a prime represents
a derivative with respect to the extra dimension $y$.

By combining Eqs. (\ref{6})-(\ref{7}) and (\ref{9}), we arrive at convenient
expressions for $\overset{.}{\phi },$ $\phi ^{\prime }$ and $V(\phi )$: 
\begin{equation}
\overset{.}{\phi ^{2}}\text{ }=2\left[ 2\dfrac{\overset{.}{a}^{2}}{a^{2}}-%
\dfrac{\overset{..}{a}}{a}+\dfrac{\overset{.}{a}}{a}\dfrac{\overset{.}{b}}{b}%
+\dfrac{\overset{.}{a}}{a}\dfrac{\overset{.}{u}}{u}+\dfrac{1}{2}\dfrac{%
\overset{.}{b}}{b}\dfrac{\overset{.}{u}}{u}-\dfrac{1}{2}\dfrac{\overset{..}{b%
}}{b}-\dfrac{\overset{..}{u}}{u}+\dfrac{\overset{.}{u}^{2}}{u^{2}}+\dfrac{k}{%
u^{2}}\right] ,  \label{11}
\end{equation}%
\begin{eqnarray}
\phi ^{^{\prime }2} &=&b^{2}\left[ \dfrac{1}{a^{2}}\left( -\dfrac{\overset{..%
}{a}}{a}-\dfrac{\overset{.}{a}^{2}}{a^{2}}-5\dfrac{\overset{.}{a}}{a}\dfrac{%
\overset{.}{u}}{u}+\dfrac{\overset{.}{a}}{a}\dfrac{\overset{.}{b}}{b}+2%
\dfrac{\overset{.}{b}}{b}\dfrac{\overset{.}{u}}{u}-2\dfrac{\overset{.}{u}^{2}%
}{u^{2}}-\dfrac{\overset{..}{u}}{u}+\dfrac{\overset{..}{b}}{b}-\dfrac{2k}{%
u^{2}}\right) \right]  \notag \\
&&-3\left( -\dfrac{a^{\prime 2}}{a^{2}}+\dfrac{a^{\prime \prime }}{a}+\dfrac{%
a^{\prime }}{a}\dfrac{b^{\prime }}{b}\right) ,  \label{12}
\end{eqnarray}%
\begin{eqnarray}
V(\phi ) &=&\dfrac{3}{2}\left[ \dfrac{1}{a^{2}}\left( \dfrac{\overset{..}{a}%
}{a}+\dfrac{\overset{.}{a}^{2}}{a^{2}}+5\dfrac{\overset{.}{a}}{a}\dfrac{%
\overset{.}{u}}{u}+\dfrac{\overset{.}{a}}{a}\dfrac{\overset{.}{b}}{b}+2%
\dfrac{\overset{.}{u}^{2}}{u^{2}}+\dfrac{\overset{..}{u}}{u}+\dfrac{\overset{%
.}{b}}{b}\dfrac{\overset{.}{u}}{u}+\dfrac{2k}{u^{2}}\right) \right.  \notag
\\
&&\left. -\dfrac{3}{b^{2}}\dfrac{a^{\prime 2}}{a^{2}}-\dfrac{1}{b^{2}}\dfrac{%
a^{\prime \prime }}{a}+\dfrac{1}{b^{2}}\dfrac{a^{\prime }}{a}\dfrac{%
b^{\prime }}{b}\right] .  \label{13}
\end{eqnarray}

In what follows we shall depict several important cases resulting from some
relevant particularizations.

\section{Time-dependent solutions}

Let us assume that $\phi (t,y)\equiv \phi (y)$. Thus the $G_{\text{ \ }%
5}^{0} $ component of the field equation becomes 
\begin{equation}
\dfrac{a^{\prime }}{a}\dfrac{\overset{.}{a}}{a}-\dfrac{\overset{.}{a}%
^{\prime }}{a}=0,  \label{14}
\end{equation}%
implying the possibility of spatial and temporal separation, i.e., $%
a(y,t)\equiv \alpha (y)\beta (t)$. Therefore, the equations (\ref{11})-(\ref%
{13}) take the form%
\begin{equation}
\dfrac{\overset{.}{\phi ^{2}}}{2}\text{ }=\left[ 2\left( \dfrac{\overset{.}{%
\beta }}{\beta }\right) ^{2}-\left( \dfrac{\overset{..}{\beta }}{\beta }%
\right) +\left( \dfrac{\overset{.}{\beta }}{\beta }\right) \left( \dfrac{%
\overset{.}{u}}{u}\right) -\left( \dfrac{\overset{..}{u}}{u}\right) +\left( 
\dfrac{\overset{.}{u}}{u}\right) ^{2}+\left( \dfrac{k}{u^{2}}\right) \right]
\equiv 0,  \label{15}
\end{equation}%
\begin{eqnarray}
\phi ^{^{\prime }2} &=&\frac{1}{\alpha ^{2}\beta ^{2}}\left[ -\left( \dfrac{%
\overset{..}{\beta }}{\beta }\right) -\left( \dfrac{\overset{.}{\beta }}{%
\beta }\right) ^{2}-5\left( \dfrac{\overset{.}{\beta }}{\beta }\right)
\left( \dfrac{\overset{.}{u}}{u}\right) -2\left( \dfrac{\overset{.}{u}}{u}%
\right) ^{2}-\left( \dfrac{\overset{..}{u}}{u}\right) -\left( \dfrac{2k}{%
u^{2}}\right) \right]  \notag \\
&&-3\left[ \left( \dfrac{\alpha ^{\prime \prime }}{\alpha }\right) -\left( 
\dfrac{\alpha ^{\prime }}{\alpha }\right) ^{2}\right] =-\dfrac{\Delta }{%
\alpha ^{2}}-3\left[ \left( \dfrac{\alpha ^{\prime \prime }}{\alpha }\right)
-\left( \dfrac{\alpha ^{\prime }}{\alpha }\right) ^{2}\right] ,  \label{16}
\end{eqnarray}%
\begin{eqnarray}
\frac{2}{3}V(\phi ) &=&\dfrac{1}{\alpha ^{2}\beta ^{2}}\left[ \left( \dfrac{%
\overset{..}{\beta }}{\beta }\right) +\left( \dfrac{\overset{.}{\beta }}{%
\beta }\right) ^{2}+5\left( \dfrac{\overset{.}{\beta }}{\beta }\right)
\left( \dfrac{\overset{.}{u}}{u}\right) +2\left( \dfrac{\overset{.}{u}}{u}%
\right) ^{2}+\left( \dfrac{\overset{..}{u}}{u}\right) +\left( \dfrac{2k}{%
u^{2}}\right) \right]  \notag \\
&&-3\left( \dfrac{\alpha ^{\prime }}{\alpha }\right) ^{2}-\left( \dfrac{%
\alpha ^{\prime \prime }}{\alpha }\right) =\dfrac{\Sigma }{\alpha ^{2}}%
-3\left( \dfrac{\alpha ^{\prime }}{\alpha }\right) ^{2}-\left( \dfrac{\alpha
^{\prime \prime }}{\alpha }\right) .  \label{17}
\end{eqnarray}

At this point we need to consider separately some different regimes of the
above equations. Afterwards, we solve the time-dependent part of the
solutions and, then, the spatial part is analyzed.

\subsection{The Case $\Delta \equiv \Sigma \equiv 0$ with $k=0$ and $k\neq 0$
\ }

Within the specifications outlined in this subsection epigraph, the
equations read 
\begin{equation}
\left( \dfrac{\overset{..}{\beta }}{\beta }\right) -2\left( \dfrac{\overset{.%
}{\beta }}{\beta }\right) ^{2}-\left( \dfrac{\overset{.}{\beta }}{\beta }%
\right) \left( \dfrac{\overset{.}{u}}{u}\right) +\left( \dfrac{\overset{..}{u%
}}{u}\right) -\left( \dfrac{\overset{.}{u}}{u}\right) ^{2}-\left( \dfrac{k}{%
u^{2}}\right) =0,  \label{18}
\end{equation}

\begin{equation}
\left( \dfrac{\overset{..}{\beta }}{\beta }\right) +\left( \dfrac{\overset{.}%
{\beta }}{\beta }\right) ^{2}+5\left( \dfrac{\overset{.}{\beta }}{\beta }%
\right) \left( \dfrac{\overset{.}{u}}{u}\right) +2\left( \dfrac{\overset{.}{u%
}}{u}\right) ^{2}+\left( \dfrac{\overset{..}{u}}{u}\right) +\left( \dfrac{2k%
}{u^{2}}\right) =-\Delta \beta ^{2}=0,  \label{19}
\end{equation}

\begin{equation}
\left( \dfrac{\overset{..}{\beta }}{\beta }\right) +\left( \dfrac{\overset{.}%
{\beta }}{\beta }\right) ^{2}+5\left( \dfrac{\overset{.}{\beta }}{\beta }%
\right) \left( \dfrac{\overset{.}{u}}{u}\right) +2\left( \dfrac{\overset{.}{u%
}}{u}\right) ^{2}+\left( \dfrac{\overset{..}{u}}{u}\right) +\left( \dfrac{2k%
}{u^{2}}\right) =\Sigma \beta ^{2}=0.  \label{20}
\end{equation}

When $k=0$, Eqs. (\ref{18})-(\ref{20}) constraint $\beta(t)$ and $u(t)$ as 
\begin{equation}
\beta (t)=\dfrac{C}{u(t)},  \label{21}
\end{equation}%
where $C$ is an arbitrary constant of integration.

It is interesting to notice that the set of equations (\ref{18})-(\ref{20})
with $k\neq 0$ and $u=1$, recover the result obtained by \cite{Ahmed}, whose
solution is 
\begin{equation}
\beta (t)=\beta _{0}e^{\pm \sqrt{-k}t}.  \label{22}
\end{equation}
On the other hand, for the case in which $k\neq 0$ e $u\neq 1$ and using the
following redefinition 
\begin{equation}
u(t)=f(t)\beta (t)  \label{23}
\end{equation}%
where $f(t)$ is an arbitrary function, we have%
\begin{equation}
\beta (t)=\dfrac{1}{\sqrt{f(t)}}\left[ C\pm \sqrt{-k}\int_{0}^{t}\frac{%
dt^{^{\prime }}}{\sqrt{f(t^{^{\prime }})}}\right] .  \label{24}
\end{equation}

\subsection{The Case $\Delta =-\Sigma $ with $k=0$ and $k\neq 0$}

Now, for the case specified here, the set of equations to be solved is 
\begin{equation}
\left( \dfrac{\overset{..}{\beta }}{\beta }\right) -2\left( \dfrac{\overset{.%
}{\beta }}{\beta }\right) ^{2}-\left( \dfrac{\overset{.}{\beta }}{\beta }%
\right) \left( \dfrac{\overset{.}{u}}{u}\right) +\left( \dfrac{\overset{..}{u%
}}{u}\right) -\left( \dfrac{\overset{.}{u}}{u}\right) ^{2}-\left( \dfrac{k}{%
u^{2}}\right) =0,  \label{25}
\end{equation}%
\begin{equation}
\left( \dfrac{\overset{..}{\beta }}{\beta }\right) +\left( \dfrac{\overset{.}%
{\beta }}{\beta }\right) ^{2}+5\left( \dfrac{\overset{.}{\beta }}{\beta }%
\right) \left( \dfrac{\overset{.}{u}}{u}\right) +2\left( \dfrac{\overset{.}{u%
}}{u}\right) ^{2}+\left( \dfrac{\overset{..}{u}}{u}\right) +\left( \dfrac{2k%
}{u^{2}}\right) =-\Delta \beta ^{2},  \label{26}
\end{equation}%
\begin{equation}
\left( \dfrac{\overset{..}{\beta }}{\beta }\right) +\left( \dfrac{\overset{.}%
{\beta }}{\beta }\right) ^{2}+5\left( \dfrac{\overset{.}{\beta }}{\beta }%
\right) \left( \dfrac{\overset{.}{u}}{u}\right) +2\left( \dfrac{\overset{.}{u%
}}{u}\right) ^{2}+\left( \dfrac{\overset{..}{u}}{u}\right) +\left( \dfrac{2k%
}{u^{2}}\right) =\Sigma \beta ^{2}.  \label{27}
\end{equation}%
All together, these equations implies $\Delta =-\Sigma $ and therefore the
solution is given by%
\begin{equation}
\beta (t)=\left[ u(t)\left( C_{1}\pm \sqrt{\dfrac{\Sigma }{3}}\int_{0}^{t}%
\frac{dt^{\prime }}{u(t^{\prime })}\right) \right] ^{-1}\text{ .}  \label{28}
\end{equation}
Note that as far as $k=0$ and $u(t)=1$, we recover result obtained in \cite%
{Ahmed}, i.e., 
\begin{equation*}
\beta (t)\propto \dfrac{1}{t},
\end{equation*}
as expected.

For $k\neq 0$, the solution is given by 
\begin{equation}
\beta (t)=-\dfrac{1}{u(t)}\sqrt{\dfrac{3}{\Sigma }}\cot \left[ C_{1}\sqrt{3k}%
-\sqrt{k}\int_{0}^{t}\dfrac{dt^{\prime }}{u(t^{\prime })}\right] \sqrt{%
k+k\tan ^{2}\left[ C_{1}\sqrt{3k}-\sqrt{k}\int_{0}^{t}\dfrac{dt^{\prime }}{%
u(t^{\prime })}\right] }.  \label{29}
\end{equation}%
Making use of the usual trigonometric relation $\sec ^{2}x=1+\tan ^{2}x$, we
have%
\begin{equation}
\beta (t)=-\dfrac{1}{u(t)}\sqrt{\dfrac{3}{\Sigma }}\cot \left[ C_{1}\sqrt{3k}%
-\sqrt{k}\int_{0}^{t}\dfrac{dt^{\prime }}{u(t^{\prime })}\right] \sqrt{k\sec
^{2}\left[ C_{1}\sqrt{3k}-\sqrt{k}\int_{0}^{t}\dfrac{dt^{\prime }}{%
u(t^{\prime })}\right] },  \label{29a}
\end{equation}%
or simply%
\begin{equation}
\beta (t)=-\dfrac{1}{u(t)}\sqrt{\dfrac{3k}{\Sigma }}\dfrac{1}{\sin \left(
C_{1}\sqrt{3k}-\sqrt{k}\int_{0}^{t}\dfrac{dt^{\prime }}{u(t^{\prime })}%
\right) }.  \label{29b}
\end{equation}%
Thus Eq. (\ref{29}) can be rewritten as%
\begin{equation}
\beta (t)=-\dfrac{1}{u(t)}\sqrt{\dfrac{3k}{\Sigma }}\sec \left( \sqrt{k}%
\int_{0}^{t}\dfrac{dt^{\prime }}{u(t^{\prime })}\right) ,  \label{29c}
\end{equation}%
where use was made of 
\begin{equation}
\pm \sin x=\pm \cos (\pi /2-x)\text{ \ \ with \ \ }x=-\sqrt{k}\int_{0}^{t}%
\dfrac{dt^{\prime }}{u(t^{\prime })}\text{ and }C_{1}\sqrt{3k}=\pi /2.
\label{29d}
\end{equation}

Note that, similarly to the case of Eq. (\ref{22}), one can reproduce the
results obtained in \cite{Ahmed}, in which the solutions given in (\ref{29c}%
), for $k=1$, $k=-1$ (both for $u=1$) are respectively 
\begin{equation}
\beta (t)\propto \sec \left( t\right) ,
\end{equation}%
and%
\begin{equation}
\beta (t)\propto \sec \text{h}\left( t\right) .
\end{equation}

We shall investigate the cosmological outputs of the obtained solutions in
Section V.

\section{Spatial-dependent solutions}

Now we shall consider the spatial part of the equations (\ref{11}), (\ref{12}%
) and (\ref{13}): 
\begin{equation}
\phi ^{\prime 2}=-\dfrac{\Delta }{\alpha ^{2}}-3\left[ \left( \dfrac{\alpha
^{\prime \prime }}{\alpha }\right) -\left( \dfrac{\alpha ^{\prime }}{\alpha }%
\right) ^{2}\right] ,  \label{30}
\end{equation}%
\begin{equation}
\dfrac{2}{3}V(\phi )=\dfrac{\Sigma }{\alpha ^{2}}-3\left( \dfrac{\alpha
^{\prime }}{\alpha }\right) ^{2}-\left( \dfrac{\alpha ^{\prime \prime }}{%
\alpha }\right) ,  \label{31}
\end{equation}%
\begin{equation}
\phi ^{\prime \prime }+4\left( \dfrac{\alpha ^{\prime }}{\alpha }\right)
\phi ^{\prime }-\dfrac{dV}{d\phi }=0.  \label{32}
\end{equation}
There are two interesting cases, concerning the separation constants which
we are going to investigate in detail. The first one is given by the
vanishing of both.

\subsection{The case $\Delta =\Sigma =0$}

For better dealing with the system of equations we use the re-definition, 
\begin{equation}
\alpha (y)=e^{A(y)}.  \label{def}
\end{equation}
In the light of (\ref{def}), the Equations (\ref{30})-(\ref{32}) become%
\begin{equation}
\dfrac{1}{3}\phi ^{\prime 2}=-A^{\prime \prime }\text{, }\dfrac{2}{3}V(\phi
)=-A^{\prime \prime }-4\left( A^{\prime }\right) ^{2},  \label{34}
\end{equation}%
and%
\begin{equation}
\phi ^{\prime \prime }+4A^{\prime }\phi ^{\prime }=\dfrac{dV}{d\phi }.
\label{36}
\end{equation}
Assuming, for ulterior convenience, that 
\begin{equation}
\phi ^{\prime }=rW_{\phi }(\phi )\text{, }A^{\prime }=sW(\phi ),  \label{37}
\end{equation}
and substituting the equations (\ref{37}) in (\ref{36}) we find $s=-1/3$.
Therefore the potential acquires the form%
\begin{equation}
V(\phi )=r^{2}\left[ \dfrac{1}{2}\left( W_{\phi }\right) ^{2}-\dfrac{2}{3}%
W(\phi )^{2}\right].  \label{39}
\end{equation}
The derivative of Eq. (\ref{39}) with respect to $\phi$, $V_{\phi}$, reads 
\begin{eqnarray}
V_{\phi } &=&r^{2}\left[ W_{\phi \phi }W_{\phi }-\dfrac{4}{3}W_{\phi }W(\phi
)\right]  \notag \\
&=&r^{2}\left[ \frac{d}{d\phi }\left( \frac{W_{\phi }^{2}}{2}\right) -\dfrac{%
4}{3}\frac{d}{d\phi }\left( \frac{W(\phi )^{2}}{2}\right) \right]  \notag \\
&=&\frac{d}{d\phi }\left[ r^{2}\left( \frac{W_{\phi }^{2}}{2}-\dfrac{2}{3}%
W(\phi )^{2}\right) \right].  \label{40}
\end{eqnarray}%
Therefore the potential itself can be expressed as 
\begin{equation}
V(\phi )=\dfrac{W_{\phi }^{2}}{2}-\dfrac{2}{3}W(\phi )^{2},  \label{41}
\end{equation}%
and the Eqs. (\ref{37}) become 
\begin{equation}
\phi ^{\prime }=W_{\phi }(\phi )\text{, }A^{\prime }=-\dfrac{1}{3}W(\phi ),
\label{42}
\end{equation}%
with%
\begin{equation}
A(y)=-\dfrac{1}{3}\int W\left[ \phi (y)\right] dy.  \label{44}
\end{equation}

Now we turn ourselves to a different arrangement of the separation constants.

\subsection{The case $\Delta =-\Sigma \neq 0$}

Making use of the expression (\ref{def}) in (\ref{30})-(\ref{32}), one gets%
\begin{equation}
\phi ^{\prime 2}=\Sigma e^{-2A}-3A^{\prime \prime }\text{, }\dfrac{2}{3}%
V(\phi )=\Sigma e^{-2A}-4\left( A^{\prime }\right) ^{2}-A^{\prime \prime },
\label{45}
\end{equation}%
and 
\begin{equation}
\phi ^{\prime \prime }+4A^{\prime }\phi ^{\prime }=\dfrac{dV}{d\phi }.
\label{47}
\end{equation}%
Combining the equations (\ref{45}), we obtain%
\begin{equation}
V(\phi )=3[A^{\prime \prime }-2\left( A^{\prime }\right) ^{2}]+\dfrac{3}{2}%
\phi ^{\prime 2}.  \label{48}
\end{equation}%
By assuming \cite{Bazeia} that 
\begin{equation}
\phi ^{\prime }=aW_{1\phi }(\phi ),\text{ }A^{\prime }=bW_{2}(\phi )\ ,
\label{49}
\end{equation}%
\begin{equation}
\ \phi ^{\prime \prime }=a^{2}W_{1\phi \phi }W_{1\phi },\ A^{\prime \prime
}=abW_{2\phi }W_{1\phi },  \label{50}
\end{equation}%
and substituting Eqs. (\ref{49}) and (\ref{50}) into (\ref{48}), we have%
\begin{equation}
V(\phi )=3[abW_{2\phi }W_{1\phi }-2b^{2}W_{2}^{2}]+\dfrac{3}{2}a^{2}W_{1\phi
}^{2}.  \label{51}
\end{equation}%
Taking the derivative of (\ref{51}) with respect to $\phi $ one gets 
\begin{equation}
\dfrac{dV}{d\phi }=3[abW_{2\phi \phi }W_{1\phi }+abW_{1\phi \phi }W_{2\phi
}-4b^{2}W_{2}W_{2\phi }]+3a^{2}W_{1\phi }W_{1\phi \phi },  \label{52}
\end{equation}%
and inserting (\ref{52}) in the scalar field equation (\ref{47}), one
arrives at the following consistence equation for the superpotential 
\begin{equation}
a[4bW_{2}-3bW_{2\phi \phi }-2aW_{1\phi \phi }]W_{1\phi }=3b[aW_{1\phi \phi
}-4bW_{2}]W_{2\phi },  \label{53}
\end{equation}%
where $a=1$ and $b=1/3$.

Now we define the following quantities%
\begin{equation}
W_{1}=W+\lambda Z\ \ \ \ \ \ \text{and}\ \ \ \ W_{2}=W+\sigma Z,  \label{54}
\end{equation}%
from which one can see that some terms can be written as a total derivative,
and the equation (\ref{53}) takes the form%
\begin{gather}
\dfrac{d}{d\phi }\left[ \dfrac{2}{3}W^{2}+\frac{1}{2}W_{\phi }^{2}-\frac{2}{3%
}\lambda \sigma Z^{2}+\frac{\alpha }{2}(2\sigma -\lambda )Z_{\phi }^{2}+%
\frac{4}{3}\sigma ^{2}Z^{2}+\sigma W_{\phi }Z_{\phi }+\frac{4}{3}\sigma WZ%
\right]  \notag \\
+\frac{4}{3}(\sigma -\lambda )WZ_{\phi }=0.  \label{55}
\end{gather}
In order to deal with a concrete and exact case, we assume that 
\begin{equation}
W=C_{1}+C_{2}Z+C_{3}Z_{\phi \phi }.  \label{56}
\end{equation}
For simplicity, some authors consider $Z(\phi )=W(\phi )$ \cite{Bazeia,Sade}%
. Here we shall consider a more general case, substituting (\ref{56}) into
Eq. (\ref{55}). This procedure leads to 
\begin{eqnarray}
&&\dfrac{2}{3}C_{1}^{2}-C_{4}+\dfrac{4}{3}C_{1}C_{3}Z_{\phi \phi }+\dfrac{2}{%
3}C_{3}Z_{\phi \phi }^{2}+\dfrac{1}{2}C_{3}^{2}Z_{\phi \phi \phi
}^{2}+Z_{\phi }Z_{\phi \phi \phi }(C_{2}C_{3}+C_{3}\beta )  \notag \\
&&+Z_{\phi }^{2}\left( \dfrac{C_{2}^{2}}{2}-\dfrac{2C_{3}\lambda }{3}-\dfrac{%
\lambda ^{2}}{2}+C_{2}\sigma +\dfrac{2}{3}C_{3}\sigma +\lambda \sigma \right)
\notag \\
&&+Z^{2}\left( \dfrac{2C_{2}^{2}}{3}-\dfrac{2C_{2}\lambda }{3}+2C_{2}\sigma -%
\dfrac{2\lambda \sigma }{3}+\dfrac{4\sigma ^{2}}{3}\right)  \notag \\
&&+Z\left[ \dfrac{4C_{1}C_{2}}{3}-\dfrac{4C_{1}\lambda }{3}+\dfrac{%
8C_{1}\sigma }{3}+Z_{\phi \phi }\left( \dfrac{4C_{2}C_{3}}{3}+\dfrac{%
4C_{3}\sigma }{3}\right) \right] =0.  \label{57}
\end{eqnarray}

In order to keep some similarity with the well known literature, we look for
a solution as the one presented in Ref. \cite{Gremm}, for instance. We
choose thereof 
\begin{equation}
Z(\phi )=Z_{0}\cos (v\phi +s).  \label{58}
\end{equation}%
In the light of Eq. (\ref{57}), gathering $Z$, $Z_{\phi }^{2}$ and $Z^{2}$
terms together, we have%
\begin{eqnarray}
&&\frac{2}{3}C_{1}^{2}-C_{4}+\dfrac{4Z}{3}\left( C_{2}-C_{3}v^{2}-\lambda
+2\sigma \right) C_{1}+Z_{\phi }^{2}\left[ \frac{C_{2}^{2}}{3}+\frac{%
C_{3}v^{4}}{2}\right.  \notag \\
&&\left. -\dfrac{2C_{3}\lambda }{3}-\dfrac{\lambda ^{2}}{2}+C_{2}\sigma +%
\frac{2C_{3}\sigma }{3}+\lambda \sigma -v^{2}(C_{2}C_{3}+C_{3}\sigma )\right]
+Z^{2}\left[ \frac{2C_{2}^{2}}{3}+\frac{2C_{3}v^{4}}{3}\right.  \notag \\
&&\left. -\frac{2C_{2}\lambda }{3}+2C_{2}\sigma -\frac{2\lambda \sigma }{3}+%
\frac{4\sigma ^{2}}{3}-v^{2}\left( \frac{4C_{2}C_{3}}{3}+\frac{4C_{3}\sigma 
}{3}\right) \right] =0.  \label{59}
\end{eqnarray}%
By taking the $Z$ coefficient equal to zero 
\begin{equation}
C_{2}-C_{3}v^{2}-\lambda +2\sigma =0,  \label{60}
\end{equation}%
we get as the solution for $v$, the following expression%
\begin{equation}
v=\sqrt{\dfrac{C_{2}-\lambda +2\sigma }{C_{3}}},  \label{61}
\end{equation}%
where $C_{1}=\pm $ $\sqrt{3/2C_{4}}$ and $C_{2}=C_{3}+\lambda -2\sigma $ if $%
v=1$, in order to full accomplish the consistence constraint. In this vein,
the expression (\ref{56}) for $W$,\ becomes%
\begin{equation}
W=C_{1}+(\lambda -2\sigma )\cos \phi ,  \label{W}
\end{equation}%
with the choice $Z_{0}=1$ and $s=0$ in equation (\ref{58}).

Using Eq. (\ref{W}) into (\ref{49}) and (\ref{50}), we obtain%
\begin{equation}
\phi ^{\prime }=(\sigma -\lambda )\sin \phi ,  \label{62}
\end{equation}%
and%
\begin{equation}
A^{\prime }=\frac{1}{3}[(\lambda -\sigma )\cos \phi ].  \label{63}
\end{equation}%
Therefore, the solutions for $\phi $ and $A$, are respectively given by%
\begin{equation}
\phi (y)=2\text{ arccot}[e^{2y(\lambda -\sigma )}],  \label{64}
\end{equation}%
and%
\begin{equation}
A(y)=\dfrac{1}{3}y(C_{1}-\lambda +\sigma )+\frac{1}{6}\ln (1+e^{4y(\lambda
-\sigma )}).  \label{65}
\end{equation}%
By means of Eqs. (\ref{64}) and (\ref{65}) for the expression (\ref{48}), we
obtain the following shape for the potential

\begin{equation}
V(\phi )=\dfrac{2(\lambda -\sigma )^{2}(1-G(\phi )[14+G(\phi )])}{3(1+G(\phi
)^{2}}
\end{equation}%
where, $G(\phi )\equiv \left[ \arctan (\phi /2)\right] ^{2(\lambda -\sigma
)^{2}}$.

In the figures (1) and (2) we depict the profiles of $\phi (y)$ and $A(y)$
in the relevant range where the scalar field is also varying. Before to
delve into the effective quantities study, we remark by passing that,
despite the rather non-trivial functional form of the obtained solutions,
the resulting spacetime is after all well behaved. In fact, all the
Kretschmann scalars associated to the solutions are finite.

\begin{figure}[tbp]
\caption{Behavior of $\protect\phi (y)$ as a function of $y$ for $\Delta
\neq \Sigma $, $\protect\lambda =2$ and $\protect\sigma =1$}\centering
\includegraphics[width=0.25\columnwidth]{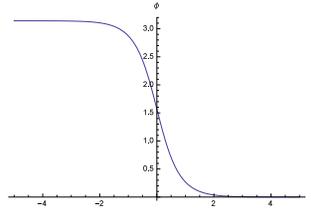}
\end{figure}

\begin{figure}[tbp]
\caption{The warp factor function $A(y)$ as a function of $y$ for $\Delta
\neq \Sigma $, $\protect\lambda =2$, $\protect\sigma =1$ and $C_{1}=0$.}%
\centering
\includegraphics[width=0.25\columnwidth]{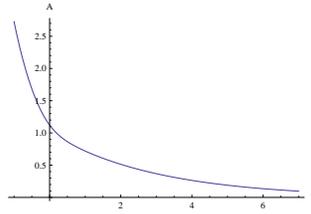}
\end{figure}

\section{Effective Hubble parameter and scale factor}

In this section, from the found solutions for $\beta (t)$, we obtain the
effective Hubble parameter as well as the effective scale factor. The Hubble
parameter, $H=\overset{.}{a}/a$, is used to measure the expansion rate of
the universe. The time elapsed in this scenario is the so called proper time
or cosmic time.

In view of the transformation 
\begin{equation}
d\tau =\beta (t)dt,  \label{66}
\end{equation}%
it is possible to construct the effective Hubble parameter 
\begin{equation}
H_{eff}=\dfrac{1}{\beta (t)}\dfrac{d}{dt}[\ln (\beta (t)u(t)]=\dfrac{1}{%
a_{eff}}\dfrac{da_{eff}}{d\tau }=\frac{\overset{.}{a}_{eff}}{a_{eff}},
\label{67}
\end{equation}%
where%
\begin{equation}
a_{eff}(t)=\beta (t)u(t).  \label{68}
\end{equation}

\subsection{The case $\Delta =\Sigma =0$}

Consider for a while the case that $k=0$, so that 
\begin{equation}
\beta (t)=\dfrac{C}{u(t)}  \label{69}
\end{equation}%
and therefore 
\begin{equation}
H_{eff}(\tau )=0\text{ \ \ \ and \ \ }a_{efe}\text{\ }=constant .  \label{70}
\end{equation}%
This case leads, then, to the typical Hubble parameter describing a static
universe, sometimes called Einstein's universe.

Within this case ($\Delta =\Sigma =0$) but now with $k\neq 0$, the solution
is given by 
\begin{equation}
\beta (t)=\dfrac{1}{\sqrt{f(t)}}\left[ C\pm \dfrac{\sqrt{-k}}{\sqrt{f(t)}}%
\int_{0}^{t}\dfrac{dt^{^{\prime }}}{\sqrt{f(t^{^{\prime }})}}\right].
\label{71}
\end{equation}%
Hence we have 
\begin{equation}
d\tau =\beta (t)dt=\dfrac{dt}{\sqrt{f(t)}}\left[ C\pm \dfrac{\sqrt{-k}}{%
\sqrt{f(t)}}\int_{0}^{t}\dfrac{dt^{^{\prime }}}{\sqrt{f(t^{^{\prime }})}}%
\right] =dz[C\pm \sqrt{-k}z],  \label{72}
\end{equation}%
where we defined that $dz=dt/\sqrt{f(t)}$. Thus, we find the expression for
the cosmic time $\tau $ as given by 
\begin{equation}
\tau =Cz\pm z^{2}\sqrt{-k}/2 .  \label{73}
\end{equation}%
Choosing, for simplicity, $C=0$ and using $f(t)\equiv e^{-2at}$, we have $%
z=\left( 1/a\right) e^{at}$ and, consequently%
\begin{equation}
\tau =\pm \frac{\sqrt{-k}}{2}\frac{1}{a^{2}}e^{2at}.  \label{74}
\end{equation}%
Therefore 
\begin{equation}
t=\frac{1}{2a}\ln \left( \pm \frac{2a^{2}\tau }{\sqrt{-k}}\right) ,
\label{75}
\end{equation}
and it is possible see that returning to Eq. (\ref{71}) one arrives at 
\begin{equation}
\beta (t)=\pm \dfrac{\sqrt{-k}}{a}e^{2at}.  \label{76}
\end{equation}%
Finally, substituting the expressions (\ref{76}) and (\ref{75}) into (\ref%
{67}), one can verify that the effective Hubble parameter decays with the
inverse of $\tau ,$while that scale factor is linearly growing and real as
far as $k<0.$ 
\begin{equation}
H_{eff}(\tau )=\dfrac{1}{\tau }\text{ \ and \ }a_{eff}(\tau )=2\sqrt{-k}%
\text{ }\tau .  \label{77}
\end{equation}
It is reasonable that the behavior of $H_{eff}(\tau )$ is the one expected
both for the matter and radiation dominated phase of the universe.
Unfortunately, however, the corresponding scale factor grows much faster
than it should in a realistic scenario.

\subsection{The case $\Delta =-\Sigma \neq 0$}

Now let us consider the case in which $\Delta =-\Sigma \neq 0$ both for $k=0$
and $k\neq 0$. Firstly, we analyze the simplest case with $k=0$, leading to 
\begin{equation}
\beta (t)=\dfrac{1}{u(t)}\dfrac{1}{C_{1}\pm \sqrt{\dfrac{\Sigma }{3}}%
\int_{0}^{t}\dfrac{dt^{\prime }}{u(t^{\prime })}},  \label{78}
\end{equation}%
and from Eq. (\ref{63}),%
\begin{equation}
d\tau =\beta (t)dt=\frac{dt}{u(t)(C_{1}\pm bz)}\rightarrow \tau =\mp \frac{1%
}{b}\ln (C_{1}\pm bz)  \label{79}
\end{equation}%
where $b=\sqrt{\Sigma /3}$ and $z=\int_{0}^{t}dt^{\prime }/u(t^{\prime })$.
Inserting Eqs. (\ref{78}) and (\ref{79}) into the expression (\ref{67}), one
gets 
\begin{eqnarray}
H_{eff}(t) &=&u(t)(C_{1}\pm bz)\frac{d}{dt}\left[ \ln \left( \dfrac{1}{%
(C_{1}\pm bz)}\right) \right] =  \notag \\
&=&u(t)(C_{1}\pm bz)\frac{d}{dt}(\pm b\tau )=u(t)(C_{1}\pm bz)\pm b\frac{1}{%
u(t)(C_{1}\pm bz)}  \notag \\
&=&\pm b,  \label{80}
\end{eqnarray}%
and therefore%
\begin{equation}
H_{eff}(\tau )=\pm \sqrt{\dfrac{\Sigma }{3}},  \label{81}
\end{equation}
whose integration leads to 
\begin{equation}
a_{eff}(\tau )=a_{0}\exp \left( \pm \sqrt{\dfrac{\Sigma }{3}}\tau \right) .
\label{83}
\end{equation}

We can note that the above solution to $a_{efe}(\tau )$ with $k=0$ is
similar to the solution found for an usual universe (without any brane), and
dominated by the vacuum energy, i.e.,%
\begin{equation}
a(t)\propto \exp \left( \pm \sqrt{\dfrac{\Lambda }{3}}t\right) .  \label{84}
\end{equation}%
where $\Lambda >0$ is the cosmological constant. This fact is more important
that it may sound. In fact, in the context of thick braneworlds, as
discussed in the Introduction, there is no how to directly relate the
four-dimensional cosmological constant with some counterpart quantity in
five dimensions, or even some property of the brane modeling. In this
approach, however, we see that the separation constant (necessary to solve
the gravitational system endowed with a brane) takes the place of the
four-dimension cosmological constant. The solution given in (\ref{83})
could, indeed, represent the current phase of accelerated expansion of our
universe in a $\Lambda CDM$ (Lambda Cold Dark Matter) model.

Finally, for the case in which $k\neq 0$, and remembering that%
\begin{equation}
\beta (t)=-\dfrac{1}{u(t)}\sqrt{\dfrac{3k}{\Sigma }}\sec \left( \sqrt{k}%
z\right) ,  \label{86}
\end{equation}%
where $z\equiv \int_{0}^{t}\dfrac{dt^{\prime }}{u(t^{\prime })}$, it can be
seen that%
\begin{equation}
d\tau =\beta (t)dt=-\sqrt{\dfrac{3k}{\Sigma }}\sec \left( \sqrt{k}z\right)
dz.  \label{87}
\end{equation}%
By taking $dz=dt/u(t)$ and integrating the above expression, we obtain%
\begin{equation}
\tau =-\sqrt{\dfrac{3k}{\Sigma }}\ln \left\vert \sec \left( \sqrt{k}z\right)
+\tan \left( \sqrt{k}z\right) \right\vert .  \label{88}
\end{equation}%
Now, by inverting the Eq. (\ref{88}) for $z$ as a function of $\tau $, one
gets 
\begin{equation}
z=\frac{2}{\sqrt{k}}\arccos \left\{ \frac{\left[ 1+\exp (\tau \sqrt{\Sigma /3%
})\right] ^{1/2}}{\sqrt{2}}\right\} .  \label{89}
\end{equation}%
The Hubble parameter, then, reads 
\begin{equation}
H_{eff}=\dfrac{1}{\beta (t)}\dfrac{d}{dt}[\ln (\beta (t)u(t)]=\dfrac{d}{%
d\tau }\left[ \ln \left( -\sqrt{\dfrac{3k}{\Sigma }}\sec \left( \sqrt{k}%
z\right) \right) \right] .  \label{90}
\end{equation}%
In order to reach Eq. (\ref{90}) we have used $d\tau =\beta (t)dt$ in the
first equality. Substituting Eq. (\ref{88}) in (\ref{89}) one arrives at%
\begin{equation}
H_{eff}(\tau )=\dfrac{d}{d\tau }\left[ \ln \left( -\sqrt{\dfrac{3k}{\Sigma }}%
\sec \left( 2\arccos \left\{ \dfrac{\left[ 1+\exp (\tau \sqrt{\Sigma /3})%
\right] ^{1/2}}{\sqrt{2}}\right\} \right) \right) \right] ,  \label{91}
\end{equation}%
and therefore%
\begin{equation}
H_{eff}(\tau )=\sqrt{\dfrac{\Sigma }{3}}\tanh \left[ \sqrt{\dfrac{\Sigma }{3}%
}\tau \right] .  \label{92}
\end{equation}%
By means of (\ref{92}) the effective scale factor is given by%
\begin{equation}
a_{eff}(\tau )=\ln \left[ \cosh \sqrt{\dfrac{\Sigma }{3}}\tau \right] .
\label{93}
\end{equation}%
The above Hubble parameter, despite the fact that it presents an expected
behavior at large values of $\tau $, is again too fast at lower values of $%
\tau $.

\section{Concluding Remarks}

We have investigated exact solutions for a FRW braneworld, whose brane is
performed by a bulk scalar field. The general idea was to find out explicit
solutions which could, at least in some regime, to perform the large scale
dynamics of the universe. In the course of our analysis a plenty of
possibilities had appeared. Among them, we believe we pay attention to the
most physically appealing cases.

In some aspects, the physical outputs can model a specific era of the known
universe, as in the case represented by Eq. (\ref{77}) in which the matter
and radiation phases can be reached. By the same token, in the specific $k=0$%
, $\Delta \neq -\Sigma $ case, the separation constant $\Sigma $ can mimic a
four-dimensional cosmological constant for a de-Sitter-like universe.
Therefore, the late-time acceleration can be modeled without regarding to
any type of dark energy.

Currently we are delving into the possibility to describe more aspects of
the cosmic evolution. To accomplish that, more bulk scalar fields, as well
as different potentials may be in order. We shall postpone these
generalizations for a future work.

\section{Acknowledgments}

ASD and JMHS thanks to CNPq for financial support. PMLTS acknowledge CAPES
for financial support.

%%%%%%%%%%%%%%%%%%%%%%%%%%%%%%%%%%%%%%%%%%%%%%%%%%%%%%%%%%%%%%%%%%%%%%%%%%%%%%%%%%%%%


\begin{thebibliography}{9}
\bibitem{RS1} L. Randall and R. Sundrum, Phys. Rev. Lett., \textbf{83}:46900
(1999).

\bibitem{RS2} L. Randall and R. Sundrum, Phys. Rev. Lett., \textbf{83}:3370
(1999).

\bibitem{Gob} M. Gogberashvili, Europhys.Lett. \textbf{49}:396 (2000).

\bibitem{Ahmed} Aqell Ahmed, Bohdan Grzadkowski, Jose Wudka, Journal of High
Energy Physics \textbf{61} (2014)

\bibitem{Bazeia} D. Bazeia, F.A. Brito and L. Losano, JHEP 11, \textbf{64}
(2006).

\bibitem{Sade} J. Sadeghi and A. Mohammadi, Eur. Phys. J. C \textbf{49}, 859
(2007).

\bibitem{Gremm} M. Gremm, Phys. Rev. D \textbf{62} (2000).

\bibitem{roldao} A. E. Bernardini, R. T. Cavalcanti, R. da Rocha, Gen.
Relat. Grav. \textbf{47} (2014).
\end{thebibliography}
\end{document}